\documentclass[11pt]{article}
\usepackage[pdftex]{hyperref,graphicx,color}
\newcommand\farcs{\mbox{$.\!\!^{\prime\prime}$}}

\newcommand{\arcsec}{^{\prime\prime}}
\def\la{\,\lower2truept\hbox{${<\atop\hbox{\raise4truept\hbox{$\sim$}}}$}\,}
\def\ga{\,\lower2truept\hbox{${>\atop\hbox{\raise4truept\hbox{$\sim$}}}$}\,}

\begin{document}
\noindent {\bf{\Large ngVLA Memo \#14}}
\begin{center}
\ \\
{\Large {\bf Short Spacing Considerations for the
    ngVLA}}\footnote{This work was done as part of the ngVLA Community Studies Program.}
\ \\
David T. Frayer (Green Bank Observatory)\\
2017 June 08
\end{center}

\begin{abstract}

  The next generation Very Large Array project (ngVLA) would represent
  a major step forward in sensitivity and resolution for radio
  astronomy, with ability to achieve 2 milli-arcsec resolution at 100
  GHz (assuming a maximum baseline of 300 km).  For science on spatial
  scales of $\ga 1\arcsec$, the ngVLA project should consider the use
  of a large single dish telescope to provide short-spacing data.
  Large single-dish telescopes are complementary to interferometers
  and are crucial to providing sensitivity to spatial scales lost by
  interferometry.  Assuming the current vision of the ngVLA (300 18m
  dishes) and by studying possible array configurations, I argue that
  a single dish with a diameter of $\ga 45$m with approximately 20
  element receiver systems would be well matched to the ngVLA for
  mapping observations.

\end{abstract}

\section{Introduction}

For interferometers there is a large trade-off between spatial
resolution and surface-brightness sensitivity.  As one moves the
interferometer dishes to longer and longer baselines to achieve higher
spatial resolution, the surface-brightness sensitivity decreases
rapidly.  The magnitude of this affect can be lessen to some degree by
tapering the interferometric data to lower resolution and increasing
the number of short baselines (e.g., [1]).  However, the inherent
trade-off between spatial resolution and brightness sensitivity is
unavoidable.  Data from a single-dish telescope could help to mitigate
the loss of surface-brightness sensitivity and recover spatial
information on scales larger than is visible with the ngVLA.  This
memo discusses recommendations for using a single dish to compensate
for the lack of ``zero-spacing'' data for the ngVLA.  The 100m Green
Bank Telescope (GBT) is used for a comparison in this study, since the
GBT is the only facility currently operating over the full range of
ngVLA frequencies (1--115\,GHz) with high sensitivity.

\section{ngVLA Compared to the VLA,  GBT, and ALMA}

The relative point source sensitivity for radio facilities is
basically the ratio of the system temperature $T_{sys}$ divided by the
total effective collecting area.  This can be expressed as
\begin{equation}
\frac{\sigma_{2}}{\sigma_{1}} =  \frac{T_{sys2}}{T_{sys1}}\frac{\eta_{a1}}{\eta_{a2}}
\left(\frac{D_{1}}{D_{2}}\right)^{2} \frac{N_{{\rm eff}1}}{N_{{\rm eff}2}},
\end{equation}
where $\eta_a$ is the aperture efficiency, $D$ is the diameter of the
telescope, and $N_{\rm eff}$ is the effective number of antennae.  For
a single-dish $N_{\rm eff}=1$, and $N_{\rm eff} =
\eta_{c}(N(N-1))^{0.5}$ for an inteferometer, where $\eta_{c}$ is the
correlator efficiency (which is 0.88 for ALMA, 0.93 for the VLA, and
is assumed to be 0.93 for the ngVLA).  Equation~(1) assumes the same
bandwidth and integration time.  For single-dish comparisons, this
equation should only be used for line sensitivities since single-dish
continuum observations are typically limited by $1/f$ noise.  The
physical collecting area of the ngVLA is about 6, 10, and 13 times
that of the VLA, GBT, and ALMA, respectively, which would provide a
major improvement in point-source sensitivity.

Figure~1 shows the relative point-source line sensitivity compared
with the GBT as a function of frequency given by equation (1).  The
values below 84 GHz for ALMA assume the predicted Band-2 performance
(when/if this band is available in the future).  The values for the
ngVLA are based on the information tabulated on the ngVLA web page and
in ngVLA memo\#5 [2].  The parameters ($\eta_a$, $T_{sys}$) estimated
for the ngVLA in the 3mm band are similar to the GBT performance and
were taken to match the GBT across the full 3mm band.  A 40m single
dish with the same performance parameters as the GBT is shown for
comparison.  The GBT has similar point-source sensitivity as the VLA
and ALMA.

The relative time to map an area to the same depth for two different
facilities is related to their relative sensitivities given in
equation (1) as
\begin{equation}
  \frac{t_{2}}{t_{1}} =\left(\frac{\sigma_{2}}{\sigma_{1}}\right)^{2} \left(\frac{D_{2}}{D_{1}}\right)^{2}. 
\end{equation}
The $(D_{2}/D_{1})^{2}$ term represents the relative number of pointings
required to cover the same area, i.e., a smaller dish has a bigger beam size
and requires fewer pointings compared to a larger dish. 

Figure~2 shows the relative point-source mapping speeds and highlights
the power of the ngVLA.  For comparison, the lines for the GBT are for
one, 10, 100, and 1000 feeds.  Large single dishes (which have small
beams) require multiple feeds to be competitive in point-source
mapping.  For example, the K-band 7-element focal-plane-array on the
GBT is well matched to the VLA, and a 100-element array at W-band
would be well matched to ALMA (currently the GBT is limited to 16
feeds in the 3mm-band using Argus).  Even with 1000 feeds on the GBT
(which is possible for the 3mm-band), the ngVLA point-source mapping
speed would be unmatched.  The ngVLA would be, by far, the fastest
point-source mapping instrument at high-frequencies compared with any
current or any other planned instrument.

Of course, not all astronomical targets are point sources at the
resolutions envisioned for the ngVLA.  For extended objects the
performance of the ngVLA depends significantly on source size.
Surface-1brightness sensitivity ($\sigma_{T}$) is typically given in
Kelvin for radio astronomy and is proportional to point-source
sensitivity divided by the beam area on the sky.  The ratio of
brightness sensitivities can be written as
\begin{equation}
  \frac{\sigma_{T2}}{\sigma_{T1}} = \frac{\sigma_{2}}{\sigma_{1}}
  \left(\frac{\theta_1}{\theta_2}\right)^2, 
\end{equation}
where $\theta$ is the angular resolution of the synthesized beam for
an interferometer and the primary beam for a single dish.  In
comparison with the compact ``D-array'' configuration of the VLA, the
ngVLA has an improvement in resolution of a factor of 300 which would
correspond to a decrease of $\sim 10^{5}$ in brightness sensitivity
for the same point-source sensitivity. 

Figure~3 shows the brightness sensitivity of the ngVLA compared with
the VLA and ALMA in their most compact configurations as well as with
the GBT and a 40m single-dish telescope as function of frequency.  The
ngVLA data points that include tapering are for a $1\arcsec$ beam
based on values given on the ngVLA web page, and do not include the
improvements for the proposed updated core configuration [1].  Without
significant tapering or without single-dish data, the ngVLA is not
useful for extended sources.  Figure~4 shows the relative extended
source mapping times.  The GBT would be many orders of magnitude
faster than the ngVLA interferometer in mapping extended sources.
Single-dish data would greatly complement ngVLA observations of
extended sources.

\section{Single-Dish Size Recommended for the ngVLA}
 
To estimate the size of the single dish needed for the ngVLA, three
considerations are discussed in the following sub-sections.  (\S3.1)
At the bare minimum, one needs overlapping UV coverage between the
single dish and the ngVLA.  (\S3.2) For quality imaging, there should
be sufficient UV coverage on short baselines to reconstruct the total
power information.  (\S3.3) There should be similar sensitivity for
the single-dish and the ngVLA data covering overlapping spatial
scales.

\subsection{Overlapping UV Coverage}

The general rule-of-thumb for combining single-dish and
interferometric data is that the size of the single dish should be at
least 1.5--2 times that of the shortest baselines in order to provide
sufficient overlapping UV coverage [3].  A simple way to look at this
is the following.  The maximum recoverable spatial scale of an
interferometer is
\begin{equation}
  \theta_{\rm MRS} \equiv \kappa \frac{\lambda}{L_{\rm min}},
\end{equation}
where $L_{\rm min}$ is the shortest baseline length, $\lambda$ is
wavelength, and $\kappa$ is a coefficient between 0.4--1.0.  The ALMA
project team [4] adopts a value of $\kappa \simeq 0.6$, while VLA
documentation [5] suggests a value of $\kappa \simeq 0.8$.  One needs
$\theta_{\rm MRS}$ to be at least the size of the single-dish beam
($\theta_{\rm SD}$) to be able to recover emission on the scale of the
beam size of the single dish.  The FWHM beam size of a single dish of
diameter $D_{\rm SD}$ is typically $\sim 1.2 \lambda/D_{\rm SD}$.
Using the performance of the GBT as a guide [6],
\begin{equation}
\theta_{\rm SD} = 1.18 \frac{\lambda}{D_{\rm SD}}
\end{equation}
is adopted for this memo.  Using equations (4\&5) and requiring
$\theta_{\rm MRS} \geq \theta_{\rm SD}$, results in
\begin{equation}
D_{\rm SD} \geq 1.18 \frac{L_{\rm min}}{\kappa}. 
\end{equation}

The appropriate value for the coefficient $\kappa$ depends on the
actual source distribution and signal-to-noise of the data.  Simple
sources with high signal-to-noise data can get by with larger values
of $\kappa$, while it is safer to use a smaller value of $\kappa$ for
more complicated sources at lower signal-to-noise.

Assuming a Gaussian source on the sky with a FWHM size of $\theta_{\rm
  FWHM}$, the normalized visibility amplitude ($f_{k}$) as a function
of projected baseline length in wavelength units ($L_{\lambda}$) is
given by [7] as
\begin{equation}
f_{k} = \exp(-3.56\,L_{\lambda}^{2}\theta_{\rm FWHM}^{2}).
\end{equation}
For a source whose size is equal to $\theta_{\rm MRS}$ and for the
shortest baseline length, $f_{k} = \exp(-3.56\,\kappa^{2})$.  For the
$\kappa = 0.6$ criterion, this corresponds to a visibility that is
0.28 times the source peak, while for $\kappa = 0.8$ the visibility
threshold is about 0.1 times the source peak.  Some authors adopt
$\theta_{\rm MRS} = \lambda/L_{\rm min} (\kappa=1)$, but in this case
less than 3\% of the source peak is recovered, and one would be
attempting to fit a Gaussian from only data within the wings of the
profile.  If one adopts a more conservative criterion corresponding to
where the visibility amplitude is half the intensity of a Gaussian
source peak, then $\kappa = 0.44$.  For this conservative case,
$D_{\rm SD} > 2.7 L_{\rm min}$.  For the ``ALMA'' convention ($\kappa
= 0.6$), $D_{\rm SD} > 2.0 L_{\rm min}$, while for the ``VLA''
convention ($\kappa=0.8$) $D_{\rm SD} > 1.5 L_{\rm min}$.

To avoid significant shadowing at low elevations, array antennae need
to be separated by more than their physical diameter.  At elevations
of $>45^{\circ}$, shadowing is avoided if $L_{\rm min} > \sqrt{2} D$.
If one uses the most compact configurations of PdBI or SMA as a guide,
the shortest implied ngVLA spacing would be about 27m ($L_{\rm min} =
1.5\times D$).  However, the ngVLA antennae could be placed in a
somewhat more compact configuration given the large number of antennae
for the ngVLA.  Shadowing could be minimized by optimizing the compact
core such that only a few antenna pairs would be affected at any
specific hour angle.  Adopting the ALMA Compact Array (ACA) as a
guide, one could expect the smallest ngVLA antenna spacing of about
23m for the 18m ngVLA antenna ($L_{\rm min} = 1.27\times D$).

For ngVLA observations over a range of hour angles at intermediate
elevations, the minimum projected baselines will typically be $\sim
18$\,m ($1.0\times D$), regardless of the exact antenna spacing.
However, to support observations near zenith and the spacing suggested
by the ACA, the minimum recommended single-dish size for the ngVLA is
about $27\rm{m}/\kappa$ (i.e., equation [6] with $L_{\rm min} =
1.27\times D$).

Studies combining single-dish data with interferometer data recommend
a single-dish size of 1.5--2\,$L_{\rm min}$ [3,8], which is consistent
with $\kappa \sim 0.8$--0.6 used here.  The minimum acceptable size is
1.5\,$L_{\rm min}$ for overlapping UV coverage, while for accurate
cross-calibration of data sets $D_{\rm SD} \sim 2\,L_{\rm min}$ is
recommended [3].  Based on these previous studies and the results
above, I adopt a value $\kappa=0.8$ to provide the minimum allowable
size for the single-dish and a value of $\kappa=0.6$ to provided the
recommended minimum size of the single dish.  Therefore to provide
sufficient overlapping UV coverage, the minimum allowable single-dish
size is 34m (for the 18m antennae of the ngVLA) and the recommended
minimum single-dish size is 45m.

\subsection{Image Quality}

The previous subsection discusses the minimum single-dish size based
on having data with overlapping UV coverage, but did not address
sensitivity or imaging quality.  For proper image reconstruction, one
needs a sufficient number of visibilities with overlapping UV coverage
to provide similar sensitivity with the single-dish data, but equally
important is sufficiently sampling the UV plane for image quality.
For long interferometric observations that use the Earth's rotation to
fill-in the UV plane, a few short baselines would suffice.  However,
assuming the ngVLA will carryout short ``snap-shot'' observations,
which could be envisioned for large mosaic mapping, it is important
that each short observation fills-in the UV plane sufficiently.  For
complex structures, one would need a large number of baselines, while
simpler regions could get by with fewer baselines.  Arrays with less
than 10 antennae (45 baselines) are generally not suitable for
snap-shot observations (unless multiple observations are taken at
different hour angles).  The 27-element VLA (315 baselines) is
sufficient for snap-shot observations.  One could make the case for
requiring somewhere between 45---315 baselines for sufficient UV
coverage for snap-shot observations.  Based on the studies of the ACA
[9], I adopt the equivalent of the ACA (12-element, 66 baselines) for
the minimum number of overlapping baselines to provide good UV
coverage when combining with single-dish data.

Requiring at least 66 baselines with overlapping UV coverage is an
additional constraint on the size of the single dish that is dependent
on the ngVLA array configuration.  Based on $\theta_{\rm MRS}$ being
greater than the single-dish beam size, the number of short-baselines
with overlapping UV coverage with a single-dish is the number of
baselines having a length less than $0.85 \kappa D_{SD}$.  This is the
definition of ``good baselines'' for different values of $\kappa$
presented in Table~1.  For the ``core'' configuration given in ngVLA
Memo\#12 (``Memo12-core''), this would suggest a recommended
single-dish size of 70m ($\kappa=0.6$) (Table~1).  The ``original''
configuration in Memo\#12 would require a single-dish size of 138m to
provide at least 66 good baselines with overlapping UV coverage (with
$\kappa=0.6$).

Additional configurations were analyzed to test how many short
baselines could be packed into a tight configuration to minimize the
size of the single-dish required to provide at least 66 good
baselines.  All configurations studied here used 300 18m antennae.  A
``Clark-Conway'' configuration was designed with 30\% of the antennae
(100) within the core arranged in a 10x10 grid with only 20m spacing
(ignoring the effects of shadowing).  This configuration is designated
``Max-core'' in Table~1.  Although this configuration is not practical
since shadowing would be significant except at high elevation, it is
included in this study as a comparison.  A more realistic possibility
would be having multiple ACA cores to provide additional short
baselines.  The configuration designated ACA-5core is based on the
Memo12-core configuration with its central 12 antennae replaced with a
configuration with antenna separations scaled to the ACA example and
having four additional scaled-ACA configurations centered at the
corners of a square with (x,y) displacements of $\pm 500$\,m from the
ngVLA array center.  This is the equivalent of having 5 copies of the
scaled-ACA configuration within the ngVLA central core.  The offsets
of $\pm 500$\,m were chosen for sensitivity on scales of $1\arcsec$ at
W-band which is based on one of the science drivers [10].  Antennae
were semi-randomly replaced from the longer baseline positions from
the Memo12-core configuration to provide the extra 48 central core
antennae for the ACA-5core configuration.  The ACA-5core configuration
is not an optimal for imaging, but is used here to help quantify the
number of overlapping short baselines available as a function of
single-dish size.

For the ACA-5core configuration, the recommended minimum single-dish
size is 50m ($\kappa=0.6$) which is significantly reduced from the 70m
dish size needed for the Memo12-core configuration.  This 50m size
limit is based on the constraint of having at least 66 good baselines
for image quality and is close to the 45m limit required for having
sufficient overlapping UV coverage.  With more antennae concentrated
in the core, additional configuration studies, and/or relaxing the
imaging constraints, the this 50m limit could probably be lowered to
the UV coverage limit of 45m.  Adopting $\kappa=0.6$, one cannot lower
the size of the single dish below 45m for 18m ngVLA antennae.  Table~1
shows the number of ``good'' overlapping baselines for different
values of $\kappa$ as a function of single-dish size for the four
ngVLA configurations studied here.

\subsection{Sensitivity}

For data combination, the sensitivity in the overlap region between an
interferometer and a single dish should be similar.  This has been
discussed previously by multiple groups in significant detail (e.g.,
[8]).  A somewhat simpler approach is adopted here.  Assuming the
single-dish telescope has similar performance parameters as the ngVLA
antennae and for the same amount of integration time, the sensitivity
is similar when $D_{\rm SD}^{2} \approx D_{\rm ngVLA}^{2}
\sqrt{2\,N_{b,eff}}$, where $N_{b,eff}$ is the ``effective'' number of
baselines for the ngVLA.  The noise in interferometer data is related
to the number of effective baselines by
\begin{equation}
\sigma^{2} = \frac{\sigma_{1}^{2}}{N_{b,eff}} = \frac{\sum w_{k}^2 \sigma_{k}^{2}}{(\sum w_{k})^2},
\end{equation}
where $\sigma_{1}$ is the typical noise for a single visibility ($k$)
and $w_{k}$ represent the weight for each visibility.  Assuming
$\sigma_{k} \approx \sigma_{1}$, then
\begin{equation}
N_{b,eff} \approx \frac{(\sum w_{k})^{2}}{\sum w_{k}^2}.
\end{equation}
Based on equation (7), the weights for the visibilities appropriate
for a Gaussian source size equal to the single-dish beam is
\begin{equation}
  w_{k} = \exp(-5.0(L_{\rm k}/D_{\rm SD})^{2}),
\end{equation}
where $L_{\rm k}$ is the baseline length.  Given the $-5$ exponent in
equation (10), long baseline visibilities provide very little signal
on the spatial scales sampled by a single dish.  The weights given by
equation (10) are also the weights associated with tapering the ngVLA
data to the resolution of the single-dish beam size.  Table~2 shows
$N_{b,eff}$ as a function of single-dish size for the four studied
ngVLA configurations.  These results are independent of wavelength.

To achieve similar sensitivities for the spatial scale defined by the
single dish, the ratio of integration time between the ngVLA and
the single dish for mapping an area is given by [11]
\begin{equation}
  \frac{t_{\rm SD}}{t_{\rm ngVLA}} =  \frac{2\,N_{b,eff}}{N_{\rm SD}}
  \left(\frac{D_{\rm ngVLA}}{D_{\rm
        SD}}\right)^{2},
\end{equation}
where $N_{\rm SD}$ is the number of feeds for the single dish.  Table~2
shows the mapping time ratio $t_{\rm SD}/t_{\rm ngVLA}$ as a function
of single-dish size assuming a single feed for the single dish.  This
value also represents the number of single-dish feeds required to have
similar mapping times as the ngVLA.  For the Memo12-core
configuration, these results imply that about 20 feeds are needed for
the single dish to provide similar mapping times.  For the ACA-5core
configuration, which has higher sensitivity on short-baselines,
about 40 feeds would be required for the single dish to match the
mapping speed of the interferometer.

The requirement of similar sensitivity between the ngVLA and the
single dish effectively defines the optimal number of feeds needed for
the single dish.  Interestingly, the number of feeds is nearly
independent of dish size.  Therefore, the choice of ngVLA
configuration will set the optimal number of feeds required for the
single dish, independent of single-dish size.  Larger single dishes
will provide deeper maps, but having matched sensitivities is
independent of the depth of the map.  The depth of the maps required
is driven by the science needs.  The next section shows the ngVLA
sensitivity as a function of source size which could be used to help
define the sensitivity requirement.

\section{Line Sensitivity as a Function of Source Size}

The line sensitivity of the ngVLA for the four different
configurations is computed using the number of effective baselines as
a function source size and configuration.  The number of effective
baselines is given by equation (9) with weights $w_k = f_k$ from
equation (7).  These weights are consistent with tapering the ngVLA
data to a resolution that matches the size of the source.  The
sensitivity is scaled from the full array point-source sensitivity
values using $\sigma \propto N_{b,eff}^{-0.5}$.  Plots 5-9 show the
results for the nominal ngVLA frequencies of 2, 10, 30, 80, and 100
GHz [2].  These plots highlight the strength of the ngVLA for small
spatial-scales $< 2\arcsec$.  For spatial scales $\ga 5\arcsec$, the
ngVLA is less sensitive than the GBT.  If the science requires high
sensitivity on the scales of $>5\arcsec$, then combining ngVLA data
with data from a large single dish (e.g., 100m GBT) will be superior
than using a smaller single dish.


\section{CASA Simulations}

CASA simulations were carried out at 90 GHz to test the photometric
accuracy of combining single-dish data with simulated ngVLA data.  The
results presented here are based on the Memo12-core configuration (as
provided by the ngVLA project for the community studies program).  The
longest baselines of the ngVLA were not used in the CASA simulations
due to limitations in simulating a large region ($100\arcsec$)
required for comparison with single-dish data with an input truth
image having a pixel size of less than $0.5 \lambda/L_{\max}$ (which
implies a $10^{5}\times10^{5}$ pixel (40GB) input truth image).

The task {\em simobserve} was used to simulate ngVLA mosaic
observations using an input truth image as the ``skymodel'' with a
total integration time of 1 hour.  The simulated ngVLA visibilities
were imaged using the task {\em clean} and simulated single-dish data
were used for the ``modelimage'' for the combination of ngVLA and
single-dish data.  The input single-dish images were made by
convolving the truth image with an appropriately sized Gaussian
representing a single dish with a size of 30m, 40m, 50m, and 100m.
The single-dish data were noiseless.  This is an idealized experiment
to test the ability to recover flux assuming perfect single-dish data.
All data were cleaned down to a similar level and the noise levels in
the output images were approximately 1--2\,mJy per beam with a beam
size of $1\farcs0 \times 0\farcs8$.

To test the ability to image regions with different sizes, the input
truth image was comprised of a $40\arcsec$ diameter circular pill-box,
a large Gaussian disk profile with a FWHM of $20\arcsec \times
5\arcsec$, and 17 small Gaussian sources with a FWHM of $0.6\arcsec$
(Fig. 10).  Two sets of simulations were carried out.  One set used
bright small sources, and for the other set of simulations, the peak
brightness of all the components were the same.

\subsection{Simulation with Bright Small Sources}

For the first set of simulations, the 17 small sources were much
brighter than the large circular pill-box and the Gaussian disk.  The
peak brightness for the small Gaussian sources was 240 \,mJy per
sq-arcsec, while the peak surface-brightness for the Gaussian disk was
96 \,mJy per sq-arcsec.  The level for the large circular pill-box
(1.9\,mJy per sq-arcsec) was only slightly above the noise in the
image.  The goal of this experiment is to test the photometric
accuracy of small sources in the presence of weaker extended emission.

Figure~11 shows the results of imaging the ngVLA data without any
single-dish data, while Figures~12--15 show the results for combining
the ngVLA data with data from a 30m, 40m, 50m, and 100m single dish,
respectively.  Without any single-dish data, the image results in a
significant negative bowl (which is expected when lacking zero-spacing
data).  Visually, even the ngVLA+30m imaging does a reasonable job at
recovering the large-scale emission within the image.  The imaging
improves marginally with data from a progressively larger single dish.

To quantify the results, aperture measurements were made on all the
images (Table~3).  Without single-dish data, the photometry in the
simulated ngVLA data is not very accurate even for the small sources.
Aperture measurements for the 10 small sources farthest from the
bright regions of the Gaussian disk were used to estimate the
photometric errors.  These errors represent the standard deviation
(scatter) of the aperture measurements for sources with the same input
flux densities.  Without including single-dish data, there was a
22.4\% photometric error in ngVLA image.  In comparison for the
ngVLA+100m image, there was an uncertainty of 4.6\%.  The empirical
measurement errors are estimated to be 3\% for individual sources.  By
removing the 3\% measurement error in quadrature, I estimate errors
due to imaging from these simulations and tabulate the results in
Table~3.  For the ngVLA+30m and ngVLA+40m combinations, the
photometric imaging errors are $>5$\%, while the ngVLA+50m and
ngVLA+100m combinations have photometric imaging errors of $\sim$4\%.

In addition to the photometric results, Table~3 shows the total
fractional flux recovered for the entire image and the Gaussian disk.
All the ngVLA+single-dish images recover the full input flux.  In
fact, for these simulations, the total output flux was slightly larger
than the input flux when measured over the same area.  This
discrepancy is not due to the single-dish data (since these were
noiseless), and may indicate that the simulation process and/or
imaging does not completely conserve surface brightness (tiny
systematic errors for a large number of pixels can add up over a large
area).

A primary result from this experiment is that single-dish data are
important for accurate photometry even if one is only interested in
small sources within complex regions.

\subsection{Simulation with Bright Large Sources}

For the second set of simulations, the large circular pill-box had the
same input surface brightness (9.6\,mJy per sq-arcsec) as the peak of
the small sources and the peak of the large Gaussian disk (Fig. 16).
This simulation highlights the importance of a large single dish.
Data from small single dishes do not have the resolution to properly
reconstruct the total power information in the central regions of the
pill-box and the Gaussian disk (Figs. 17--21).  In this case, the
simulation using the ngVLA+100m single-dish data is significantly
superior than smaller single-dish combinations.

Based on the theoretical arguments in Section 3.2, the Memo12-core
configuration only had a sufficient number of ``good'' baselines for
the 100m combination in these simulations (i.e., needed a 70m dish to
achieve the desired 66 ``good'' baselines).  To test the relative
importance of the size of the single-dish versus the number of short
baselines for the ngVLA, this simulation was repeated with the
Core-5ACA configuration which has five times as many short baselines.
The additional short baselines from the Core-5ACA configuration did
not improve the results in this simulation.  Therefore, the larger the
single dish, the better the results will be for recovering complex
emission from large sources when combining with ngVLA observations.

\section{Discussion}

Of the three criteria given in Section~3 for choosing the size of the
single dish, the requirement of sufficient overlapping UV coverage
appears to be the most important (Section 3.1).  The requirement of
having similar sensitivities (Section 3.3) does not constrain the
single-dish size, but rather sets the number of optimal feeds required
for the single dish.  The simulation results imply that the metric of
having 66 ``good'' baselines for image quality (Section 3.2) is not a
firm requirement.  The many 1000's of baselines available from the
ngVLA with low weights and lengths larger than the ``good'' threshold
level ($0.85 \kappa D_{SD}$) appear to provide adequate information
for image reconstruction.  For example, in the simulations presented
here, a 40m single dish performs fairly well, but only has 23 ``good''
baselines even with the relaxed definition of $\kappa=0.8$ for the
simulated Memo12-core configuration.

As derived in Section 3.1, the required single-dish size for the ngVLA
is $1.50\,D_{\rm ngVLA}/\kappa$.  For $D_{\rm ngVLA} = 18$\,m, this
corresponds to 45\,m for $\kappa = 0.6$ (recommended value) and 34\,m
(minimum value) for $\kappa = 0.8$.

The simulations show that having a larger single dish produces better
results, especially for imaging large complex sources.  Whatever the
single-dish size, it is important to have quality single-dish data.
The simulations presented here assumed noiseless single-dish data, but
previous detailed studies suggest signal-to-noise ratios of S/N$>20$
are needed to avoid significant imaging errors [8].

It would be reasonable for the ngVLA project to deploy a dedicated
single dish to be used with the array.  Many projects would benefit
from the combination of the array data with data taken with a single
dish of size 34--50m.  For extended-source projects that require
higher sensitivity (e.g., sub-mJy 1hr, 10 km/s sensitivity at 3mm),
the GBT could be used to provide the single-dish data.

The choice of single-dish properties depends on the ngVLA array
configuration as well as the ngVLA antenna size.  The core of the
ngVLA should be optimized to minimize shadowing while maximizing the
number of baselines overlapping with the single dish.  A larger single
dish provides more flexibility.  If the ngVLA needs to support
snap-shot observations, then multiple compact cores may be required to
provide a sufficient number of instantaneous short baselines for
combination with single-dish data.  In this memo, 66 ``good''
baselines was adopted as a criterion based on the ACA, but the value
of 66 as well as the definition of ``good'' is admittedly somewhat
arbitrary.  More detailed simulations should be done to test the
limits for short snap-shot integrations for large mosaic mapping with
the ngVLA.

\section{Concluding Remarks}

Assuming that the ngVLA antenna locations are not re-configurable, the
final ngVLA configuration decision will need to be a compromise
between good high-resolution imaging and the number of antennae that
will comprise the core(s).  The ngVLA project should include a large
single dish which would complement the interferometric data.  A larger
single dish relaxes the requirements on the configuration of the
compact core compared to a smaller single dish, as well as providing
better data products.  The smallest recommended single-dish size is
45m for the 18m ngVLA antenna size.  Adopting the Memo12-core
configuration suggests that the single dish should be deployed with
about 20 feeds to provide matched mapping times with the ngVLA
interferometric observations.

\section{References}

\begin{enumerate}

\item Carilli, C. L. 2016, The Strength of the Core, ngVLA Memo \#12

\item Carilli, C. L., et al. 2015, Science Working Groups Project
  Overview, ngVLA Memo \#5

\item Stanimirovic, S. 2002, Short-Spacings Correction from the
  Single-Dish Perspective, Single-Dish Radio Astronomy: Techniques and
  Applications, ASP Conference Proceedings, Vol. 278., 375-396

\item ALMA Technical Handbook, 2017 Cycle 5:
https://almascience.nrao.edu/documents-and-tools/cycle5/alma-technical-handbook/view

\item Perley, R. A. 1995, Very Large Array Observational Status Summary

\item Frayer, D.T. 2017, The GBT Beam Shape at 109 GHz, GBT Memo \#296

\item Wilner, D.J. \& Welch, W. J. 1994, ApJ, 427, 898

\item Kurono, Y., Morita, K. \& Kamazaki, T. 2009, PASJ, 61, 873

\item Wright, M.C.H. 2003, Heterogenous Imaging with the ALMA Compact
  Array, ALMA Memo \#450 

\item Leory, A. K. et al. 2015, Science Working Group 2: ``Galaxy
  Ecosystems'': The Matter Cycle in and Around Galaxies, ngVLA Memo
  \#7

\item Mason, B.S. \& Brogan, C. 2013, Relative Integration Times for
  the ALMA Cycle 1 12-m, 7-m, and Total Power Arrays, ALMA Memo \#598

\end{enumerate}

\clearpage
\begin{table}
\footnotesize
  \caption{Number of Good Overlapping Baselines}
  \begin{tabular}{c|rrrr|rrrr|rrrr|rrrr}
    \hline \hline
SD&\multicolumn{4}{c}{Memo12-orig}&\multicolumn{4}{c}{Memo12-core}&\multicolumn{4}{c}{Max-core}&\multicolumn{4}{c}{ACA-5core}\\  
Size&\multicolumn{16}{c}{for $\kappa$ values = 0.44, 0.6, 0.8, and 1.0}\\ 
(m)&0.44&0.6&0.8&1.0&0.44&0.6&0.8&1.0&0.44&0.6&0.8&1.0&0.44&0.6&0.8&1.0\\
\hline
 30&  0&  0&  2& 12&   0&   0&   4&  19&   0&   0& 180& 180&   0&   0&   1&  72\\
 35&  0&  0&  5& 15&   0&   0&   9&  37&   0&   0& 180& 342&   0&   0&  40&  86\\
 40&  0&  2& 14& 17&   0&   4&  23&  57&   0& 180& 180& 342&   0&   1&  75& 130\\
 45&  0&  3& 17& 21&   0&   7&  38&  77&   0& 180& 342& 342&   0&  15&  87& 182\\
 50&  0& 12& 17& 24&   0&  19&  57&  99&   0& 180& 342& 502&   1&  72& 130& 215\\
 55&  2& 14& 20& 26&   4&  30&  77& 132& 180& 180& 342& 790&   1&  79& 177& 261\\
 60&  3& 17& 23& 35&   6&  38&  91& 153& 180& 342& 502& 790&   3&  87& 206& 306\\
 65&  5& 17& 25& 44&  10&  54& 112& 178& 180& 342& 502& 790&  41& 108& 229& 352\\
 70& 14& 17& 26& 53&  20&  67& 134& 211& 180& 342& 790& 918&  72& 155& 269& 407\\
 75& 14& 21& 35& 57&  30&  77& 153& 253& 180& 342& 790&1310&  79& 182& 306& 461\\
 80& 15& 23& 42& 63&  38&  91& 172& 285& 342& 502& 790&1310&  87& 206& 337& 502\\
 85& 17& 24& 49& 70&  47& 104& 199& 328& 342& 502& 918&1310&  96& 219& 382& 555\\
 90& 17& 25& 54& 83&  56& 126& 225& 366& 342& 790&1058&1534& 125& 242& 421& 600\\
 95& 17& 26& 59& 88&  66& 140& 261& 416& 342& 790&1310&1654& 149& 282& 470& 670\\
100& 20& 35& 63& 95&  77& 153& 285& 462& 342& 790&1310&1870& 177& 306& 502& 722\\
\hline
\end{tabular}
\end{table}

\normalsize
%
\begin{table}
  \caption{Single Dish Comparisons with ngVLA Configurations}
  \begin{tabular}{c|rr|rr|rr|rr}
    \hline \hline
SD&\multicolumn{2}{c}{Memo12-orig}
 &\multicolumn{2}{c}{Memo12-core}
 &\multicolumn{2}{c}{Max-core}
 &\multicolumn{2}{c}{ACA-5core}\\
Size &Effective&$t_{\rm SD}/t_{\rm I}$&Effective&$t_{\rm SD}/t_{\rm I}$&Effective&$t_{\rm SD}/t_{\rm I}$&Effective&$t_{\rm SD}/t_{\rm I}$ \\
(m) &Baselines&  & Baselines&  &Baselines&  &Baselines& \\
\hline
 30& 10.8&  7.8&  23.2& 16.7&  215.8&155.4&   76.8& 55.3\\
 35& 13.4&  7.1&  34.1& 18.0&  245.3&129.7&   94.7& 50.1\\
 40& 15.4&  6.2&  46.1& 18.7&  279.9&113.4&  114.3& 46.3\\
 45& 17.2&  5.5&  59.0& 18.9&  320.4&102.5&  135.9& 43.5\\
 50& 19.2&  5.0&  72.9& 18.9&  367.6& 95.3&  159.4& 41.3\\
 55& 21.4&  4.6&  87.6& 18.8&  421.6& 90.3&  184.2& 39.5\\
 60& 24.0&  4.3& 103.4& 18.6&  481.9& 86.7&  210.3& 37.9\\
 65& 26.9&  4.1& 120.2& 18.4&  547.7& 84.0&  237.4& 36.4\\
 70& 30.2&  4.0& 138.3& 18.3&  618.3& 81.8&  265.5& 35.1\\
 75& 33.9&  3.9& 157.6& 18.2&  692.9& 79.8&  294.5& 33.9\\
 80& 37.8&  3.8& 178.1& 18.0&  770.9& 78.1&  324.3& 32.8\\
 85& 42.0&  3.8& 199.9& 17.9&  852.0& 76.4&  354.8& 31.8\\
 90& 46.5&  3.7& 223.0& 17.8&  935.7& 74.9&  386.0& 30.9\\
 95& 51.1&  3.7& 247.3& 17.8& 1021.5& 73.3&  417.7& 30.0\\
100& 56.0&  3.6& 272.8& 17.7& 1109.4& 71.9&  449.9& 29.2\\
\hline
 \end{tabular}
\end{table}

\clearpage

\begin{table}
\caption{Simulation Results for Bright Small Sources}
\begin{tabular}{lrrrrr}
    \hline \hline
Measurement & ngVLA & ngVLA+ & ngVLA+ & ngVLA+ & ngVLA+\\
            &       &  30m  &   40m    &   50m  & 100m\\
\hline
Photometric errors$^{a}$ &22.4\% & 8.3\% & 6.2\%& 5.3\%& 4.6\%\\
Imaging errors$^{b}$ &22.2\% & 7.7\% & 5.4\%& 4.4\%& 3.5\%\\
Gaussian disk flux ratio$^{c}$ & 0.76 & 1.02&1.05&1.06& 1.02\\ 
Total flux ratio$^{c}$&0.32 & 1.07&1.12&1.13& 1.06\\ 
\hline
\end{tabular}

 (a) Errors representing the scatter of photometric measurements for the
 small sources.  (b) Inferred imaging errors after removing the
 empirical measurement uncertainties.   (c) Flux ratio is measured/input.
\end{table}


\begin{table}
\caption{Simulation Results for Bright Large Sources}
\begin{tabular}{lrrrrr}
  \hline \hline
  Measurement & ngVLA & ngVLA+ & ngVLA+ & ngVLA+ & ngVLA+\\
  &       &  30m  &   40m    &   50m  & 100m\\
  \hline
  Gaussian disk flux ratio$^{a}$ & 0.22 & 0.69 &0.78&0.83& 0.92\\
  Total flux ratio$^{a}$&0.08 & 0.80 & 0.88 & 0.90& 0.95\\ 
  \hline
 \end{tabular}

 (a) Flux ratio is measured/input.
\end{table}

\clearpage
\begin{figure}[tbp]
\centering
\includegraphics[width=0.7\textwidth]{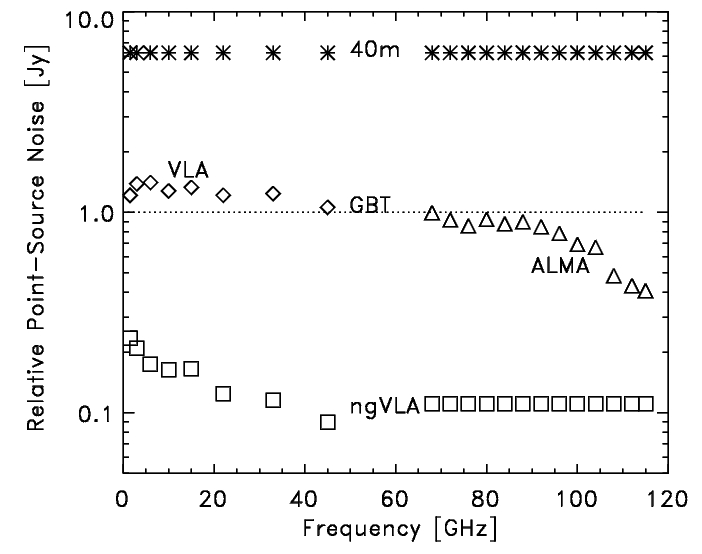}
\caption{The relative point-source line sensitivities compared to the
  GBT.  The VLA data points are given as diamonds, ALMA values are
  shown as triangles, while the ngVLA data points are shown as
  squares.  Values for a 40m single-dish with the same performance
  properties as the GBT are shown as asterisks for comparison.}
\end{figure}

\begin{figure}[tbp]
\centering
\includegraphics[width=0.7\textwidth]{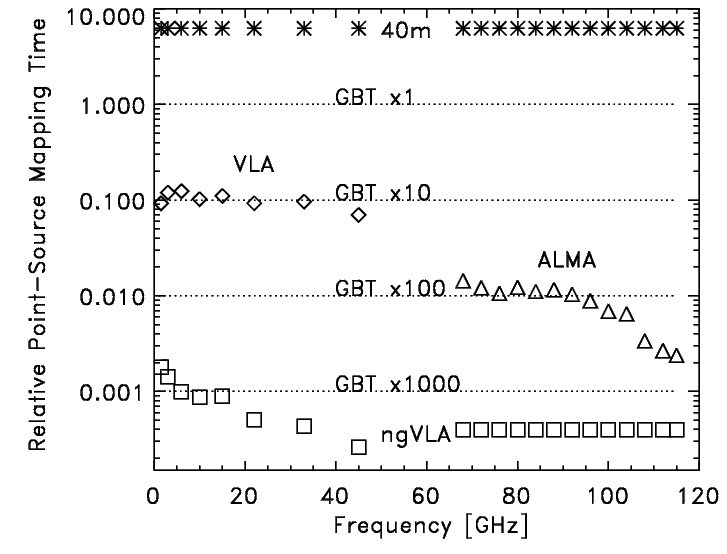}
\caption{The relative point-source mapping times compared to the GBT
  with 1, 10, 100, and 1000 feeds.  The VLA, ALMA, ngVLA, and 40m
  symbols are the same as Figure~1.  Large single dishes require
  multiple feeds to be competitive in point-source mapping.}
\end{figure}

\clearpage
\begin{figure}[tbp]
\centering
\includegraphics[width=0.68\textwidth]{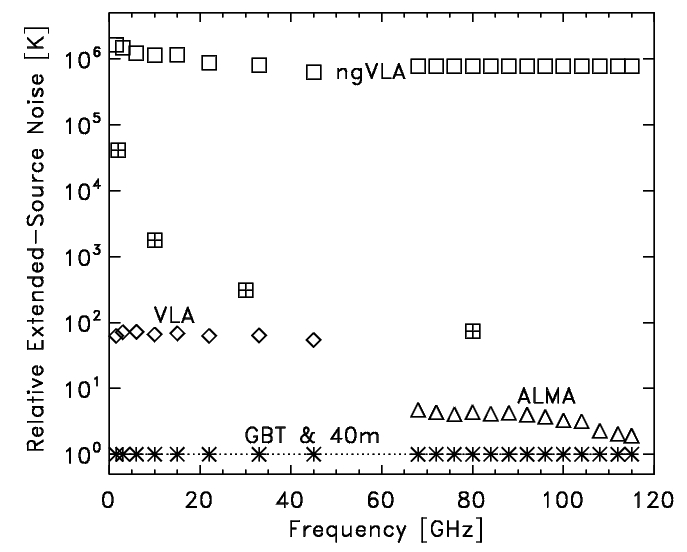}
\caption{The relative extended-source line sensitivity compared to the
  GBT.  The VLA, ALMA, ngVLA, and 40m symbols are the same as
  Figure~1.  The data for the VLA and ALMA are for their most compact
  configurations.  The squares with ``plus'' symbols represent ngVLA
  data tapered to $1\arcsec$ resolution for a sub-set of frequencies.}
\end{figure}

\begin{figure}[tbp]
\centering
\includegraphics[width=0.68\textwidth]{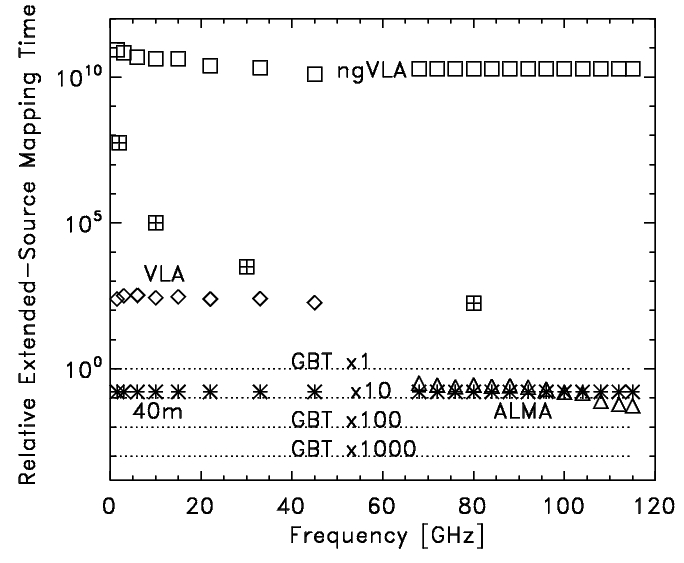}
\caption{The relative extended-source mapping times compared to the
  GBT with 1, 10, 100, and 1000 feeds.  The VLA, ALMA, ngVLA, and 40m
  symbols are the same as Figure~1\&3.  This highlights the need for
  single-dish data with the ngVLA.}
\end{figure}

\clearpage
\begin{figure}[tbp]
\centering
\includegraphics[width=0.7\textwidth]{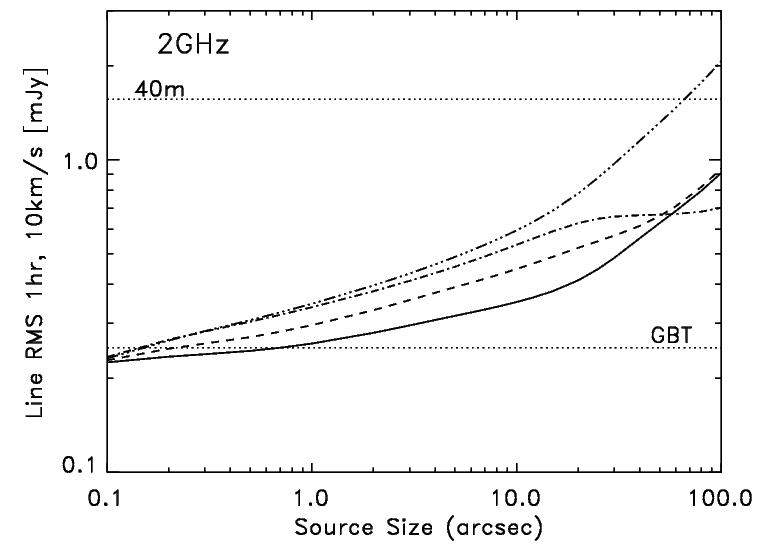}
\caption{The line sensitivity of the ngVLA for the 4 different
  configurations at 2 GHz as a function of source size.  The number of
  effective baselines decreases with increasing source size using
  weights to taper the resolution of the ngVLA data to match the size
  of the source.  The solid line is the ACA-5core configuration, the
  dashed line is the Memo12-core configuration, the dashed
  triple-dotted line is the Memo12-orig configuration, and the
  dashed-dotted line is the Maxcore configuration.  For comparison,
  the sensitivity levels of the 100m GBT and a 40m single dish are
  shown as dotted lines.}
\end{figure}

\begin{figure}[tbp]
\centering
\includegraphics[width=0.7\textwidth]{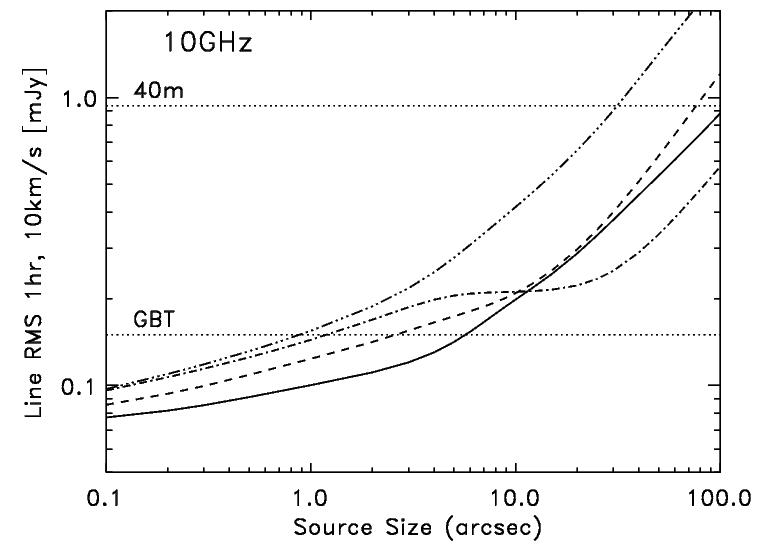}
\caption{The line sensitivity of the ngVLA for the 4 different
  configurations at 10 GHz as a function of source size.  The lines
  descriptions are the same as given in Figure~5.}
\end{figure}

\clearpage
\begin{figure}[tbp]
\centering

\includegraphics[width=0.7\textwidth]{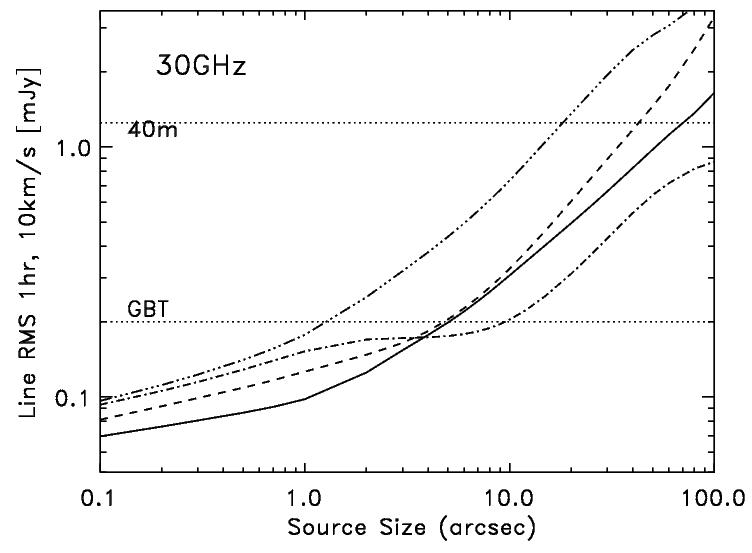}
\caption{The line sensitivity of the ngVLA for the 4 different configurations
  at 30 GHz as a function of source size.   The lines descriptions are
  the same as
  given in Figure~5.}
\end{figure}

\begin{figure}[tbp]
\centering
\includegraphics[width=0.7\textwidth]{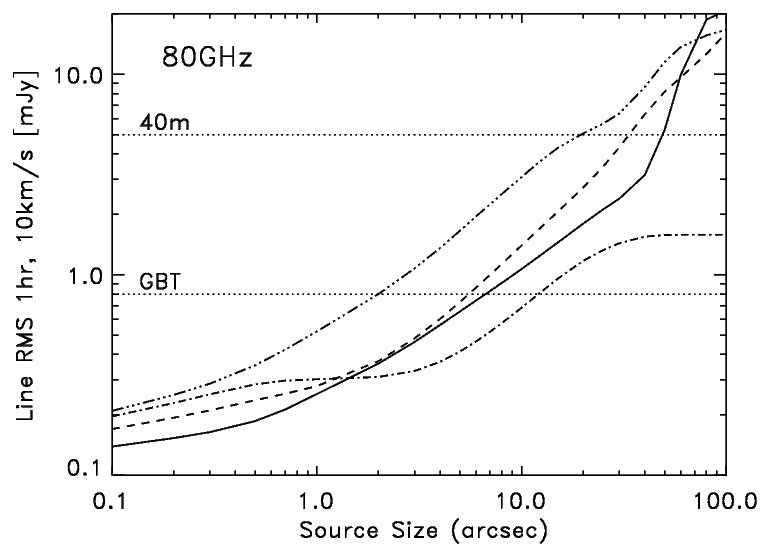}
\caption{The line sensitivity of the ngVLA for the 4 different
  configurations at 80 GHz as a function of source size.  The lines
  descriptions are the same as given in Figure~5.}
\end{figure}

\clearpage
\begin{figure}[tbp]
\centering
\includegraphics[width=0.7\textwidth]{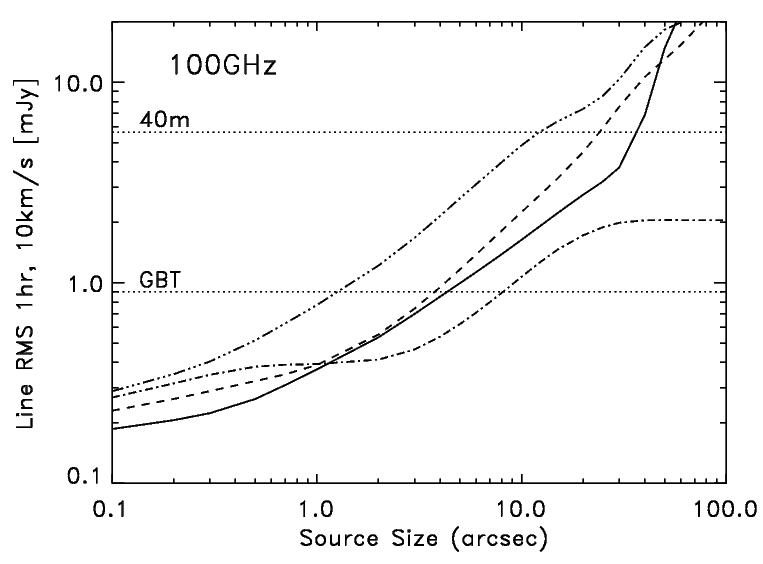}
\caption{The line sensitivity of the ngVLA for the 4 different
  configurations at 100 GHz as a function of source size.  The lines
  descriptions are the same as given in Figure~5.}
\end{figure}

\clearpage
\begin{figure}[tbp]
\centering
\includegraphics[width=0.5\textwidth]{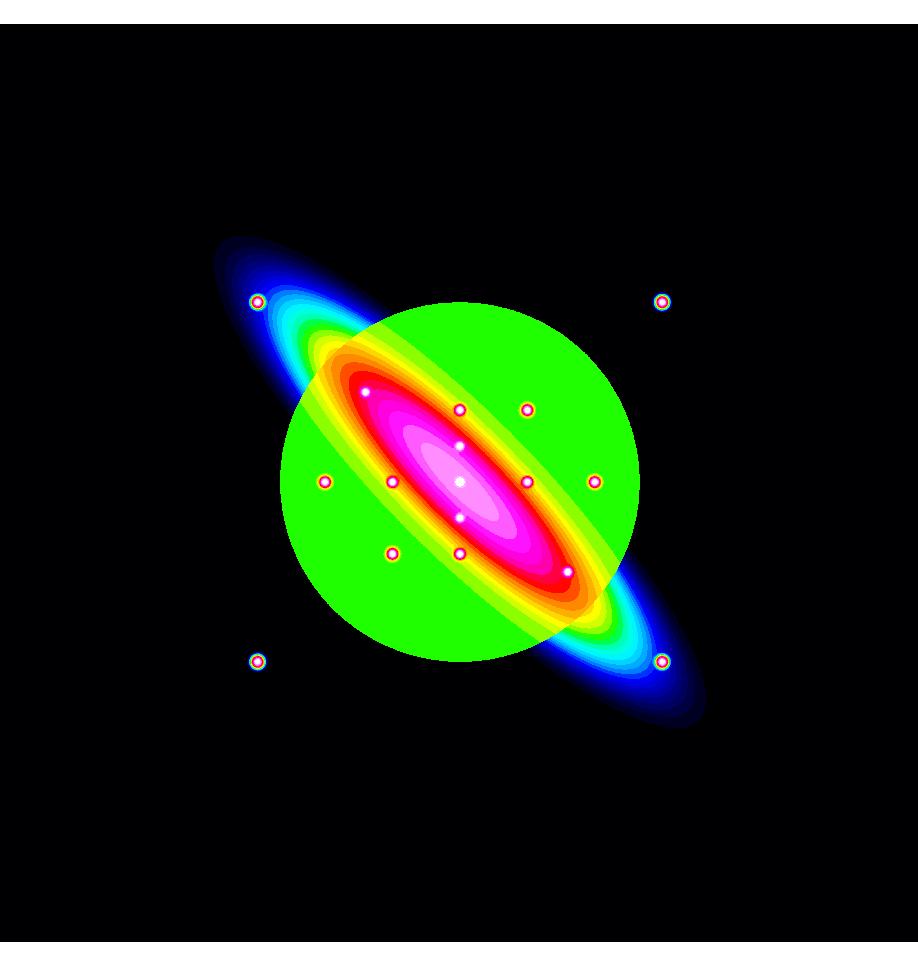}
\caption{The input truth image of the sky brightness used in the
  ``Bright Small Source'' CASA simulations.  The circular pill-box
  (green) has a $40\arcsec$ diameter which is centered on the
  $20\arcsec \times 5\arcsec$ Gaussian disk profile.  The 17 small
  sources have a $0\farcs6$ FWHM Gaussian profile and all have the
  same input flux density.  The simulation area is 102.4 arcsec on a
  side.  The background has a surface brightness of 0 mJy/sq-arcsec
  (black), and the maximum surface brightness is 340 mJy/sq-arcsec for
  the central peak.}

\end{figure}

\begin{figure}[tbp]
\centering
\includegraphics[width=0.5\textwidth]{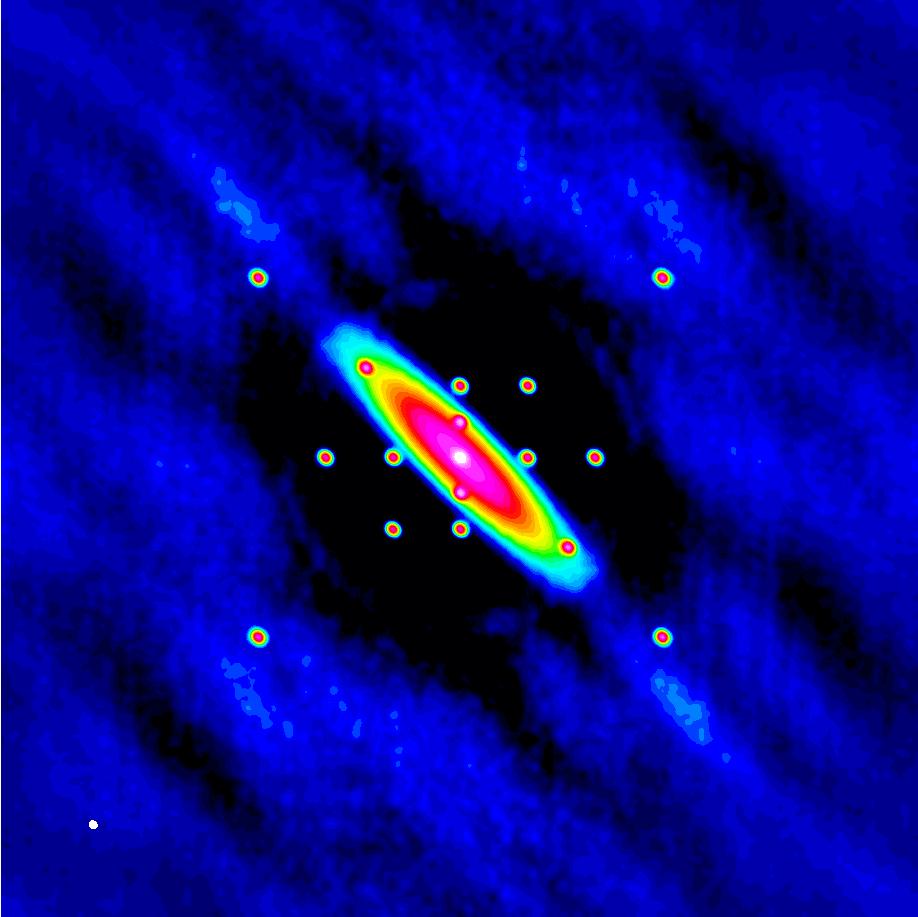}
\caption{The simulated ngVLA image without including any single-dish
  data.  Only 32\% of the total flux is recovered after cleaning.  The
white dot in the lower left shows the synthesized beam size of the
resulting image ($1\farcs0 \times 0\farcs8$). }
 \end{figure}

\clearpage
\begin{figure}[tbp]
\centering
\includegraphics[width=0.6\textwidth]{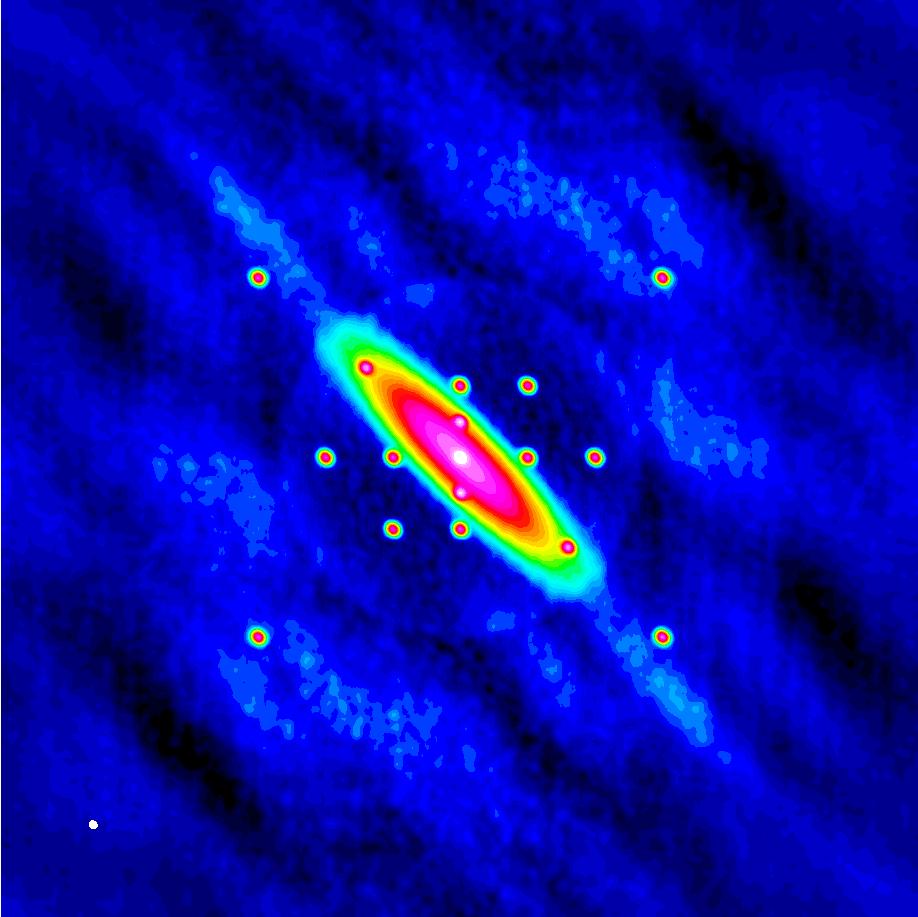}
\caption{The simulated ngVLA + 30m image.  Adding single-dish data
  fills-in the central hole.  The imaging artifacts decrease with
  progressively larger single-dish data. }
 \end{figure}

\begin{figure}[tbp]
\centering
\includegraphics[width=0.6\textwidth]{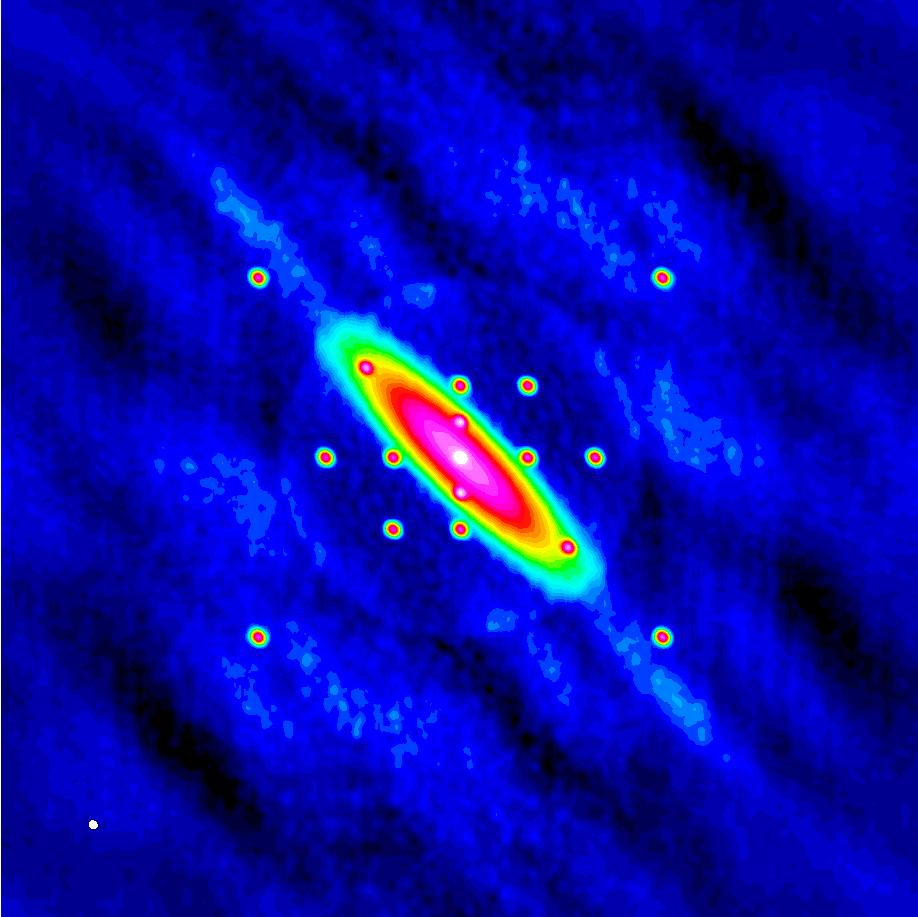}
\caption{The simulated ngVLA + 40m image.}
\end{figure}

\clearpage
\begin{figure}[tbp]
\centering
\includegraphics[width=0.6\textwidth]{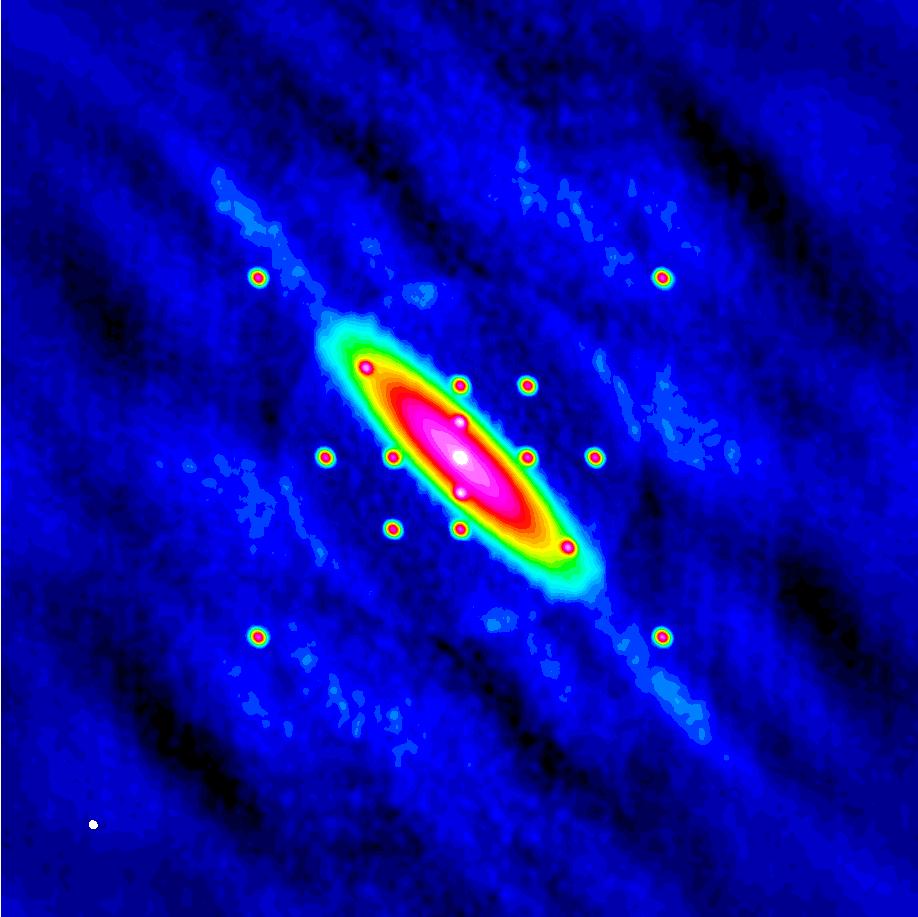}
\caption{The simulated ngVLA + 50m image.}
\end{figure}

\begin{figure}[tbp]
\centering
\includegraphics[width=0.6\textwidth]{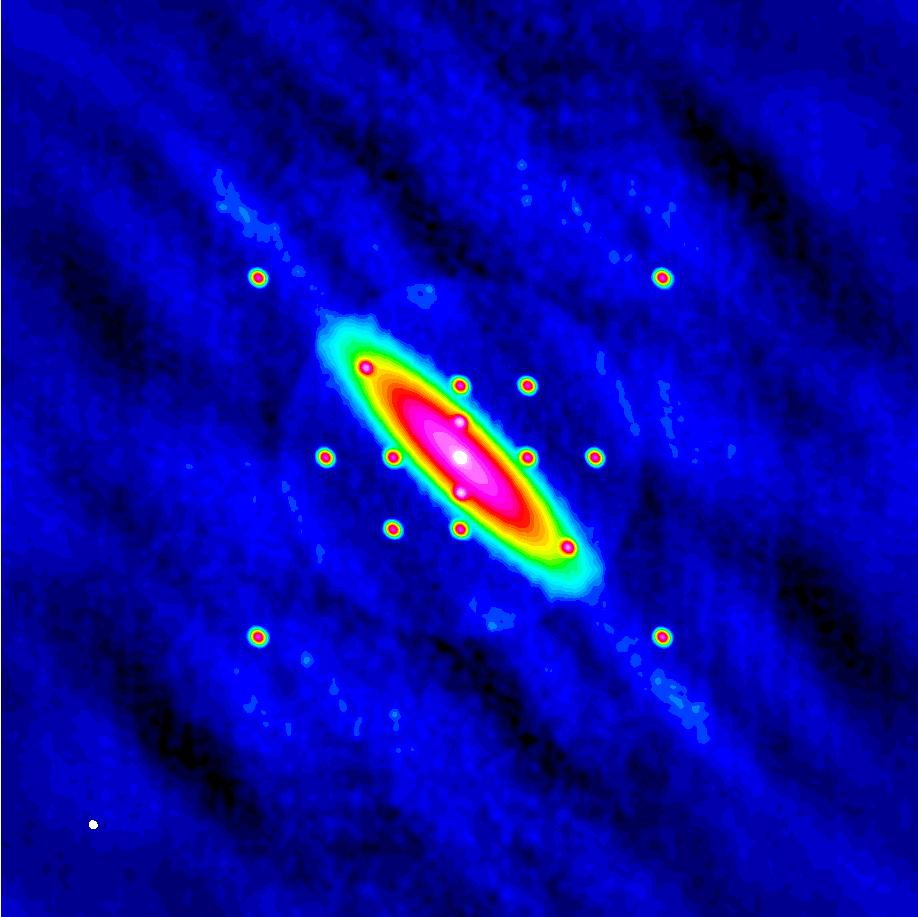}
\caption{The simulated ngVLA + 100m image.}
\end{figure}

\clearpage
\begin{figure}[tbp]
\centering
\includegraphics[width=0.52\textwidth]{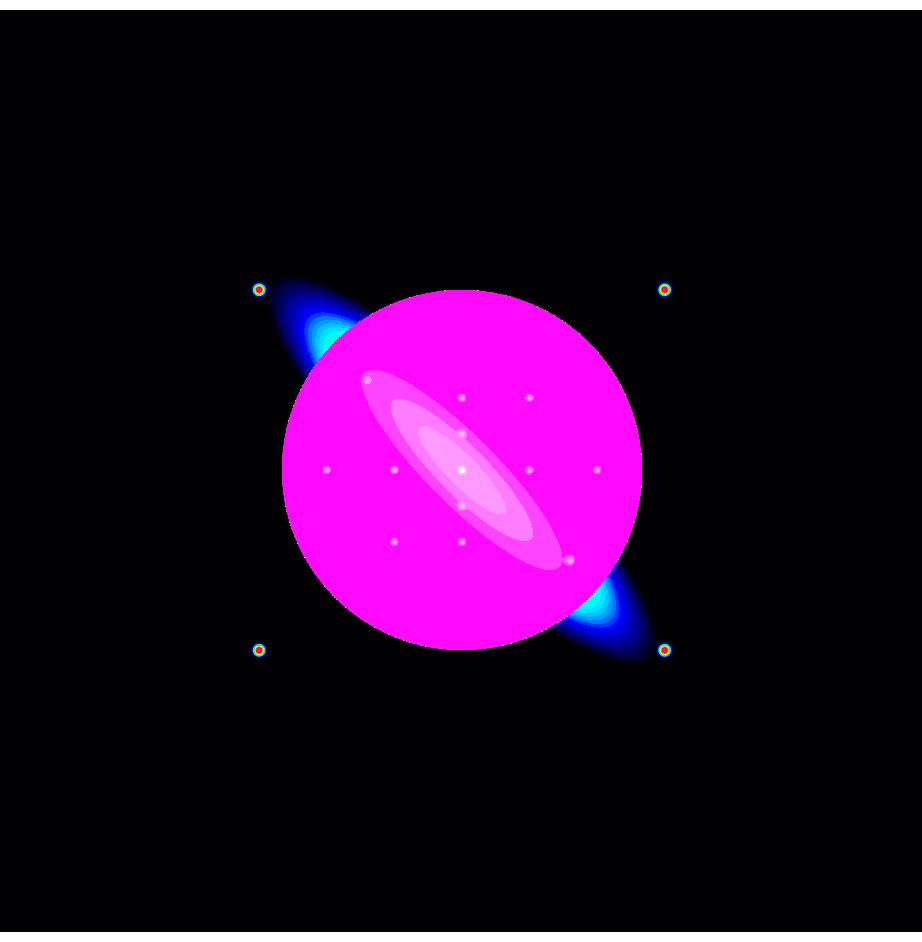}
\caption{The input truth image of the sky brightness used in the
  ``Bright Large Source'' simulations.  All components (17 small
  sources, Gaussian disk, and large circular pill-box) have the same
  peak surface-brightness of 9.6\,mJy per sq-arcsec, and most of the
  flux density is associated with the large circular pill-box.}
\end{figure}

\begin{figure}[tbp]
\centering
\includegraphics[width=0.52\textwidth]{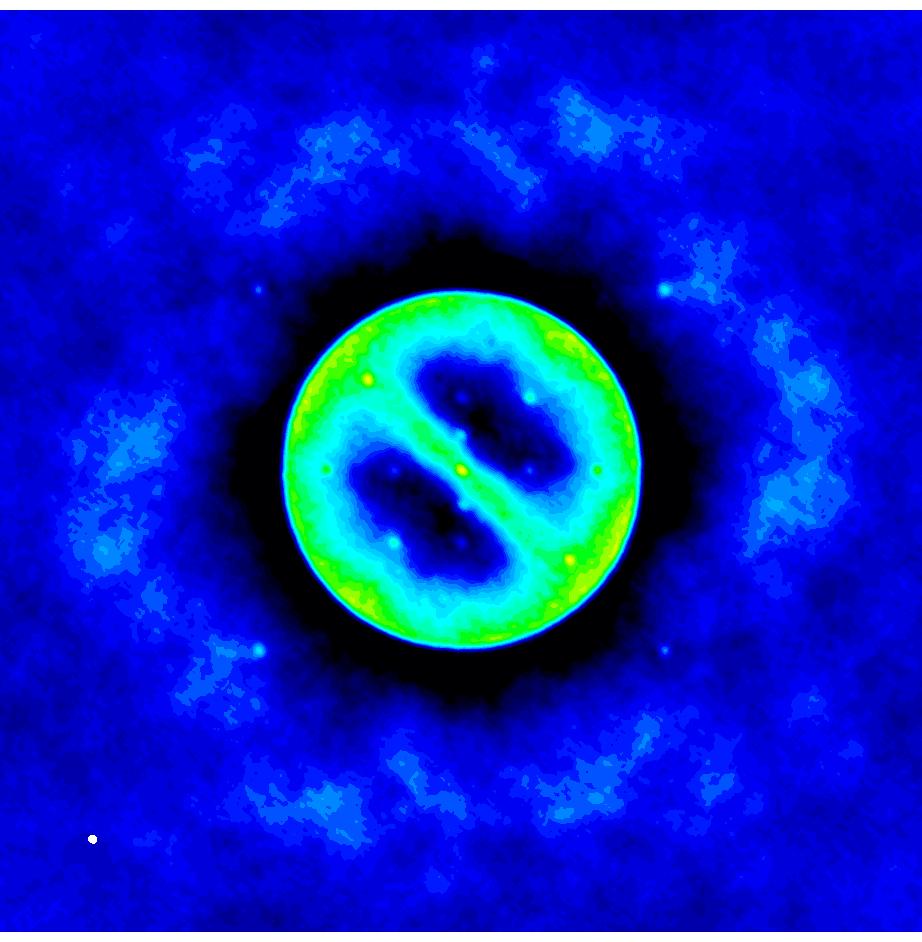}
\caption{The simulated ngVLA image without including any single-dish
  data.  Only 8\% of the total flux is recovered after cleaning.  The
  white dot in the lower left shows the synthesized beam size of the
  resulting image ($1\farcs0 \times 0\farcs8$). }
 \end{figure}

\clearpage
\begin{figure}[tbp]
\centering
\includegraphics[width=0.6\textwidth]{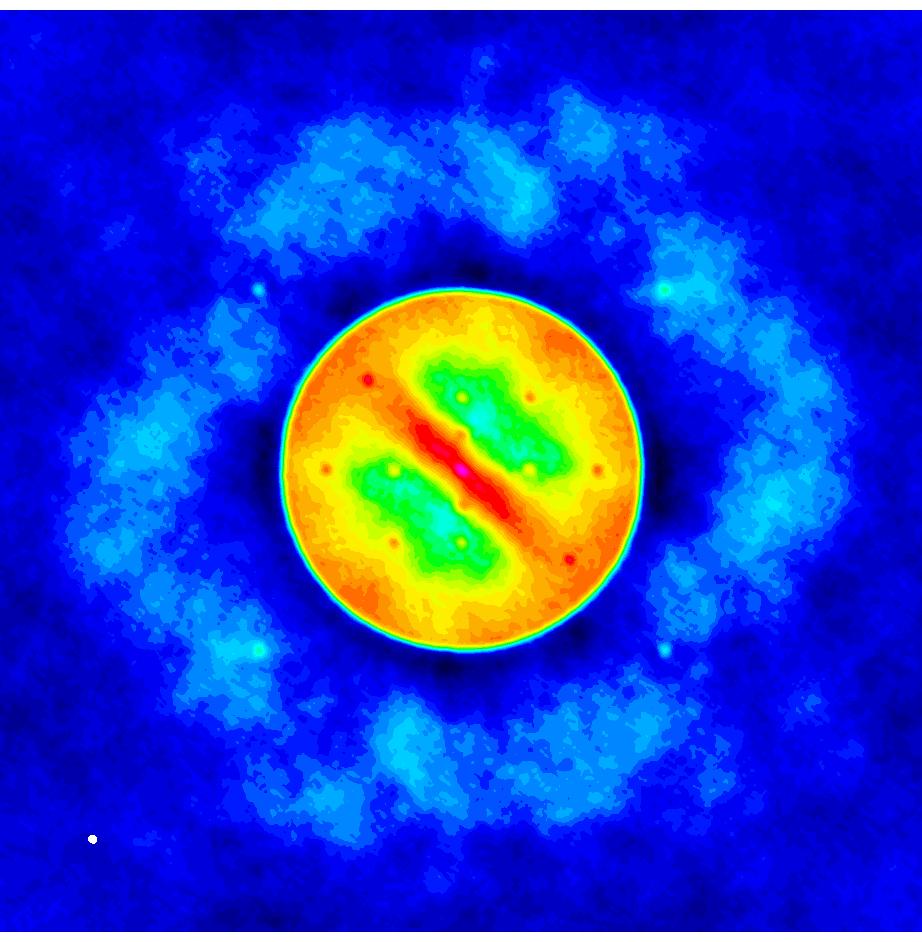}
\caption{The simulated ngVLA + 30m image. The imaging improves with
  progressively larger single-dish data. }
 \end{figure}

\begin{figure}[tbp]
\centering
\includegraphics[width=0.6\textwidth]{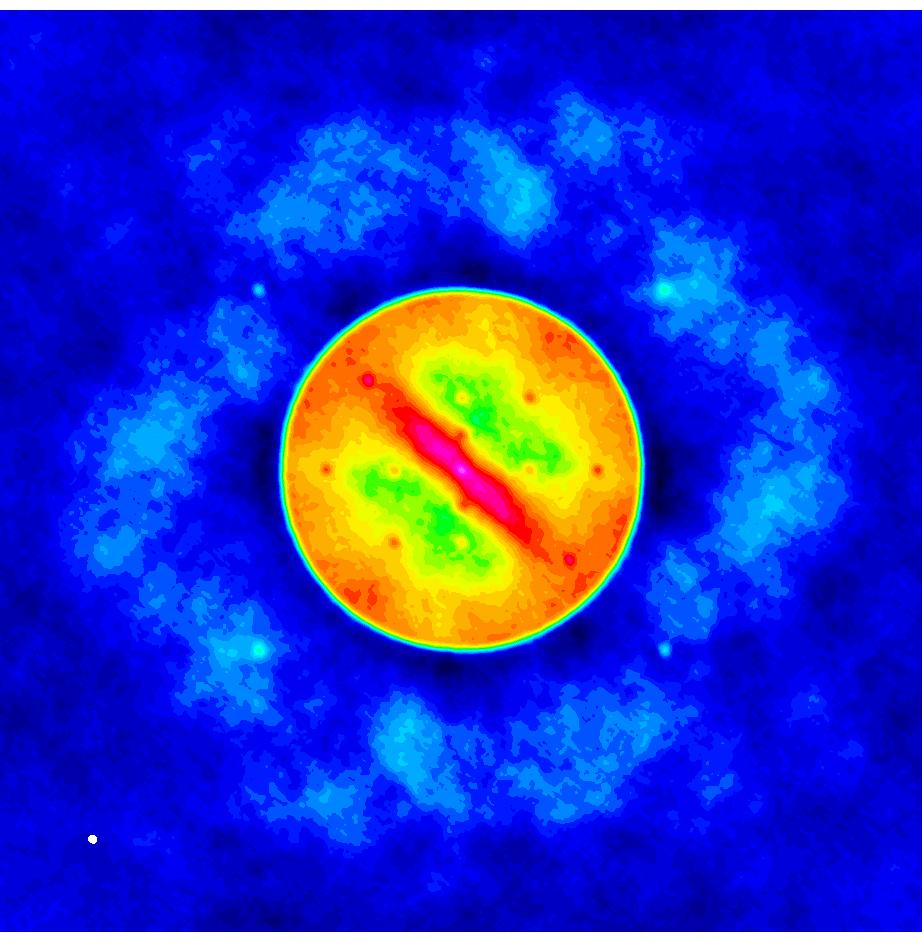}
\caption{The simulated ngVLA + 40m image.}
\end{figure}

\clearpage
\begin{figure}[tbp]
\centering
\includegraphics[width=0.6\textwidth]{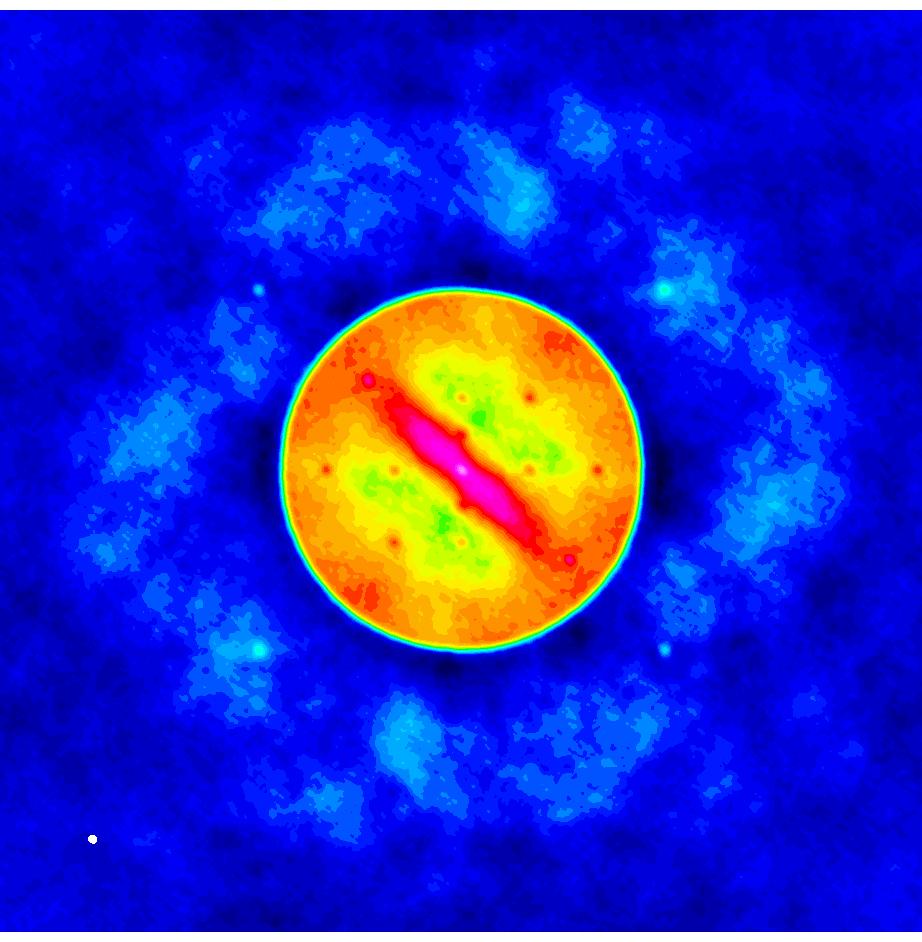}
\caption{The simulated ngVLA + 50m image.}
\end{figure}

\begin{figure}[tbp]
\centering
\includegraphics[width=0.6\textwidth]{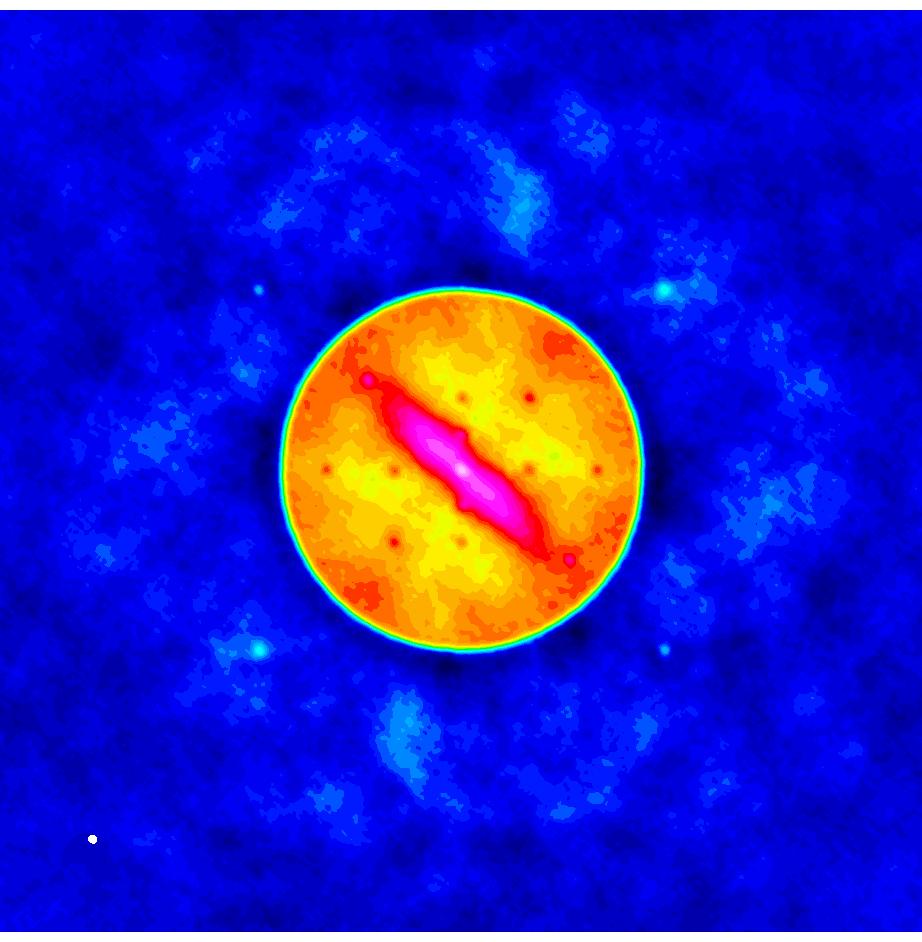}
\caption{The simulated ngVLA + 100m image.}
\end{figure}

\end{document}